\documentclass[aip,graphicx,amsmath,amssymb,preprint]{revtex4-1}
\usepackage{lineno,hyperref}
\usepackage[normalem]{ulem}
\usepackage{graphics}
\usepackage{amsmath,bm}
\usepackage{amssymb}
\usepackage{graphicx}
\usepackage{bm}
\usepackage{amsfonts}
\usepackage{color}
\usepackage{epstopdf}
\usepackage{booktabs}
\usepackage{array}
\usepackage{hyperref}
\usepackage{graphics}
\usepackage{epsfig}
\usepackage{bm}
\usepackage{epic,eepic}
\usepackage{epsfig}
\usepackage{pifont}

\usepackage{subfigure}

\usepackage{xspace}
\usepackage{graphicx,color}
\usepackage{amsmath,amssymb,amsfonts,mathrsfs}
\usepackage{url}
\usepackage[normalem]{ulem}
\usepackage{booktabs}
\usepackage{float}

\usepackage[toc,page]{appendix}
\usepackage[capitalize]{cleveref}
\usepackage{mathtools}
\usepackage[normalem]{ulem}
\usepackage{gensymb}
\usepackage{algorithm}
\usepackage{algorithmic}

\begin{document}

\title{Synthetic turbulence generator for the wall-modeled LES lattice Boltzmann method}





\author{Xiao Xue}
\email{xiaox@chalmers.se}
\affiliation{ 
Division of Fluid Dynamics, Department of Mechanics and Maritime Sciences, Chalmers University of Technology, 41296, Gothenburg, Sweden
}
\author{Hua-Dong Yao}

\author{Lars Davidson}%



\date{\today}

\begin{abstract}
The synthetic turbulence generator (STG) lies at the interface of the Reynolds averaged Navier-Stokes (RANS) simulation and large eddy simulation (LES). This paper presents a STG for the multiple-relaxation-time(MRT) lattice Boltzmann method(LBM) framework at high friction Reynolds numbers, with consideration of near wall modeling. The Reichardt wall law, in combination with a force-based method, is used to model the near wall field. The STG wall-modeled(STG-WM) LES results are compared with turbulent channel flow simulations at $Re_{\tau}=1000,2000,5200$ at different resolutions. The results demonstrate good agreement with DNS, with the adaptation length of $6$ to $8$ boundary layer thickness. This method has a wide range of potentials for hybrid RANS/LES-LBM related applications at high friction Reynolds numbers.
\end{abstract}


\maketitle


\section{INTRODUCTION}
In industrial applications related to high Reynolds number flows, wall-bounded turbulence plays a crucial role in designing aircraft, cars, and wind farms and so on (\cite{porte2011large}). Direct numerical simulations (DNS) can accurately quantify the physics by solving the Navier-Stokes equations using high-order numerical approximations and grid refinement techniques (\cite{chapman1979computational,choi2012grid}). However, DNS is computationally too expensive for real-world applications, making it impractical for use in the design cycle. Large eddy simulations (LES) can reduce grid requirements by modeling subgrid-scale eddy viscosity (\cite{Smagorinsky1963, germano1991dynamic, Sagaut2002}). Nevertheless, LES still requires fine grid resolution in the near-wall region.
requiring grid points to be proportional to 
$O(Re^{n})$ with $n=13/7$ (\cite{chapman1979computational, choi2012grid, yang2021grid}). Wall-resolved LES (WRLES) is still far from an engineering tool.
To improve computational efficiency, studies have attempted to model the near-wall region by solving Reynolds-averaged Navier-Stokes (RANS) equations and using LES in the far field, or by modeling the near-wall region with relatively few grids by reconstructing near-wall velocities. 
 In past decades, the lattice Boltzmann method (LBM) has gained popularity for simulating fluid dynamic problems at a variety of scales, from micro-nano scales (\cite{xue2020brownian, xue2018effects, xue2021lattice, chiappini2018ligament, chiappini2019hydrodynamic}) to macroscopic scales (\cite{hou1994lattice, toschi2009lagrangian, karlin1999perfect, lallemand2000theory, shao2022near}) at low Mach numbers. LBM offers an alternative to traditional methods by solving the Boltzmann equation at a mesoscopic level, instead of directly solving the Navier-Stokes equations (\cite{succi2001lattice, lallemand2000theory, kruger2017lattice, lallemand2021lattice}). LBM's parallelization-friendly nature makes it attractive due to the local update of discrete particle distribution functions. \cite{hou1994lattice} introduced the effective turbulent viscosity to model LES in LBM (LES-LBM) framework, enabling the simulation of high Reynolds number flows with increased stability. This approach combines the advantages of LES techniques with the computational efficiency of LBM.

To further improve computational efficiency, the hybrid RANS-LES approach is becoming increasingly popular as it balances accuracy and computational cost. The RANS method is used to conserve macroscopic quantities in computationally less-demanding regions, while LES provides detailed flow information in computationally-intensive areas. Generating high-quality turbulence at the RANS/LES interface is critical for achieving accurate results, as highlighted by \cite{wu2017inflow}. In the conventional CFD, studies have conducted wide range approaches by precursory DNS/LES data (\cite{schluter2004large}), velocity field "recycling" (\cite{lund1998generation, spalart2006direct, shur2011rapid}), synthetic eddy method (SEM) (\cite{mathey2006assessment, jarrin2009reconstruction, skillen2016accuracy}), or involving control techniques (\cite{roidl2011zonal, roidl2012zonal}) to generate turbulence. Most of the existing studies suffer from either relatively long turbulent-develop adaptation length, high computational cost or hard to generalize for complex geometries. \cite{shur2014synthetic} successfully developed a synthetic turbulent generator (STG) for the detached eddy simulation (DES) (\cite{shur2008hybrid}) . The STG method creates velocity fluctuations based on the Fourier coefficients which are given by the energy spectrum, whereas, for the SEM method, coherent structures are simulated by superimposing artificial eddies at the inlet plane. The results show both fast and robust with adaption length of 2-4 boundary-layer widths.

Despite it is an active area in the conventional CFD, studies using a  turbulent generator in the LBM framework are still rare. \cite{koda2015lattice} generated turbulence for the channel flow by placing a "recycled" channel flow before the inlet, which requires a pre-simulated periodic turbulent channel flow. Thus it limits its applications due to redundant preparation work.\cite{buffa2021lattice} reconstructed turbulence by using the SEM method in LBM (SEM-LBM); however, this method may suffer from relatively long adaptation length (\cite{fan2021source}). \cite{xue2022synthetic} integrated the synthetic turbulent generator (STG) in the LBM framework at $Re_{\tau} = 180$ with wall-resolved LES-LBM simulation with the adaptation length of 2-4 boundary-layer widths. However, the friction Reynolds number is relatively low and the Bhatnagar-Gross-Kroog (BGK) collision operator limited its applications for high Reynolds number cases. To reduce computational time and maintain STG accuracy, modeling the near-wall flow field with a wall-model in the LES based LBM (WMLBM) is a non-straightforward task. The first WM-LBM was proposed by \cite{malaspinas2014wall} where they successfully reconstructed the first-layer near wall velocity with the Musker wall function or log-law (\cite{musker1979explicit}). Then, follow-up works came up with the idea of reconstructing the velocity field or modeling the velocity bounce back \citep{malaspinas2014wall, haussmann2019large, maeyama2020unsteady, wilhelm2021new, asmuth2021wall}. However, the force-based method is rarely mentioned and rarely described in detail in the LBM framework.

In the present work, a synthetic turbulent generator is developed, integrated with a multiple-relaxation time (MRT) collision operator and wall-modeled LES-LBM to tackle high friction Reynolds numbers. The near-wall region is modeled via a force-based wall model, using a wall law by Reichardt (\cite{reichardt1951vollstandige}). The performance of the STG method is examined at three different friction Reynolds numbers: $Re_{\tau} = 1000$, $Re_{\tau} = 2000$, and $Re_{\tau} = 5200$ at various resolutions, and compared with DNS data from \cite{hoyas2006scaling, lee2015direct}.


\section{METHODOLOGY}\label{sec:methodology}

\subsection{The multiple-relaxation time lattice Boltzmann method}\label{sec:method-lbm}
In this work, we utilize a three-dimensional (3D) lattice model with 19 discretized directions known as the D3Q19 model. The lattice cell is located at position $\mathbf{x}$ and time $t$, with a discretized velocity set $\mathbf{c}_i$ for $i \in {0, 1, ..., Q-1}$ ($Q=19$):

\begin{equation}
\label{eq:vel-set}
\begin{aligned}
\mathbf{c}_i =  \{&(0, -1, -1), (-1, 0, -1), (0, 0, -1), (1, 0, -1), (0, 1, -1), (-1, -1, 0), \\
    &(0, -1, 0), (1, -1, 0),  (-1, 0, 0),  (0, 0, 0),  (1, 0, 0),  (-1, 1, 0), (0, 1, 0),  \\
   &(1, 1, 0), (0, -1, 1),  (-1, 0, 1),  (0, 0, 1),  (1, 0, 1),  (0, 1, 1) \}.
   \end{aligned}
\end{equation}
The weight for the discretized directions is defined as
\begin{equation}
w_{9}=\frac{1}{3}, \quad
w_{2, 6, 8, 10, 12, 16} = \frac{1}{18}, \quad
w_{0, 1, 3, 4, 5 ,7, 11, 13, 14, 15, 17, 18} = \frac{1}{36}.
\end{equation}
The evolution equation for the distribution functions, accounting for collision and forcing, can be expressed as:
\begin{equation}
\label{eq:lbe}
\mathbf{f}(\mathbf{x}+\mathbf{c}_{i}\Delta t,t+\Delta t) =\mathbf{f}(\mathbf{x}, t) - \Omega \left[\mathbf{f}(\mathbf{x},t )-\mathbf{f}^{\mbox{ eq}}(\mathbf{x},t )\right] + \mathbf{F}(\mathbf{x}, t) \Delta t,
\end{equation}
where $\Omega$ is a collision kernel and $\Delta t$ is the lattice Boltzmann time step which is set to unity. In this work, MRT collision kernel is chosen due to its higher numerical stability compared to the BGK model at high Reynolds numbers (\cite{d2002multiple}). 
\begin{equation}
\label{eq:collision}
\Omega = \mathbf{M}^{-1}\mathbf{S}\mathbf{M},
\end{equation}
where $\mathbf{M}$ is the transformation matrix from population space to moment space obtained via the Gram-Schmidt approach ($\mathbf{M}$ matrix is described in Appendix \ref{sec:A.1}). $\mathbf{S}$ is the diagonal matrix with relaxation frequencies at different moments, $\mathbf{S}=diag\{\omega_0, \omega_1, ... , \omega_{Q-1}\}$, MRT collision operator will be equivalent to BGK with $\omega_i$ set to the same value $\omega$. The frequency $\omega_i$ is the inverse of the relaxation time $\tau_i$. Note that, we set $\tau_{k} = \tau_{9}=\tau_{11}=\tau_{13}=\tau_{14}=\tau_{15}$, which are related to the kinematic viscosity $\nu$, which is 
\begin{equation}
\label{eq:nu}
\nu = c_s^2\left(\tau_{k}-\frac{1}{2}\right)\Delta t,
\end{equation}
with $c_s$ is the speed of the sound, and $c_s^2$ is equal to $1/3$ lattice Boltzmann unit (LBU). Other relaxation parameters can be found in Eq. (8,2) - Eq.(8.6). Instead of colliding in the population space, the MRT collides in the moment space, thus \cref{eq:lbe} can be rewritten as:
\begin{equation}
\label{eq:lbe_mrt}
\mathbf{f}(\mathbf{x}+\mathbf{c}_{i}\Delta t,t+\Delta t) =\mathbf{f}(\mathbf{x}, t) - \mathbf{M}^{-1}\mathbf{S} \left[\mathbf{m}(\mathbf{x},t )-\mathbf{m}^{\mbox{ eq}}(\mathbf{x},t )\right] + \mathbf{F}(\mathbf{x}, t) \Delta t,\\
\end{equation}
where $\mathbf{m}$ is the moment space component which is defined as
\begin{equation}
\label{eq:mom_f}
\mathbf{m} (\mathbf{x},t ) = \mathbf{M}\mathbf{f}(\mathbf{x},t ),
\end{equation}
the moment space equilibrium $\mathbf{m}^{\mbox{ eq}}(\mathbf{x},t )$ can be defined as
\begin{equation}
\label{eq:mom_eq}
\begin{aligned}
&\mathbf{m}_0^{\mbox{ eq}} = \rho, 												& \mathbf{m}_1^{\mbox{ eq}} &= -11 \rho  + 19 \rho (u_x^2+u_y^2+u_z^2)\\
&\mathbf{m}_2^{\mbox{ eq}} = \frac{11}{2}\rho\left(3- (u_x^2+u_y^2+u_z^2)\right), 			& \mathbf{m}_3^{\mbox{ eq}} &= \rho u_x\\
&\mathbf{m}_4^{\mbox{ eq}} = -\frac{3}{2}\rho u_x, 									& \mathbf{m}_5^{\mbox{ eq}} &= \rho u_y \\
&\mathbf{m}_6^{\mbox{ eq}} = -\frac{3}{2}\rho u_y, 									& \mathbf{m}_7^{\mbox{ eq}} &= \rho u_z\\
&\mathbf{m}_8^{\mbox{ eq}} = -\frac{3}{2}\rho u_z, 									& \mathbf{m}_9^{\mbox{ eq}}& = 2 \rho u_x^2 - \rho u_y^2 - \rho u_z^2\\
&\mathbf{m}_{10}^{\mbox{ eq}} = -\rho u_x^2 +\frac{1}{2} \rho u_y^2 +\frac{1}{2} \rho u_z^2, 	& \mathbf{m}_{11}^{\mbox{ eq}}& = \rho u_y^2 - \rho u_z^2 \\
&\mathbf{m}_{12}^{\mbox{ eq}} = -\frac{1}{2}\rho u_y^2 + \frac{1}{2}\rho u_z^2, 				& \mathbf{m}_{13}^{\mbox{ eq}} &= \rho u_x u_y\\
&\mathbf{m}_{14}^{\mbox{ eq}} = \rho u_y u_z, 										& \mathbf{m}_{15}^{\mbox{ eq}} &= \rho u_x u_z\\
&\mathbf{m}_{16}^{\mbox{ eq}} = \mathbf{m}_{17}^{\mbox{ eq}} = \mathbf{m}_{18}^{\mbox{ eq}} = 0.
\end{aligned}
\end{equation}
$\mathbf{F}(\mathbf{x}, t)$ in~\cref{eq:lbe_mrt} is the vector of $F_i(\mathbf{x}, t)$ which is the volume force acting on the fluid cell (\cite{guo2002discrete}):
\begin{equation}
\label{eq: bodyforce}
F_i(\mathbf{x}, t) = (1 - \frac{\omega_i}{2}) w_i \left [ \frac{\textbf{c}_i - \textbf{u}\left(\textbf{x},t\right)}{c_s^2} + \frac{\textbf{c}_i \cdot \textbf{u}\left(\textbf{x},t\right)}{c_s^4}\textbf{c}_i \right] \cdot \mathbf{g},
\end{equation}
where $\mathbf{g}$ is the volume acceleration. The macro-scale quantities for the density, momentum, and momentum flux tensors, can be calculated from the distribution function, the discrete velocities, and the volume force:
\begin{equation}\label{eq:density}
\rho(\mathbf{x}, t) = \sum_{i=0}^{Q-1} f_i(\mathbf{x}, t), \\
\end{equation} 
\begin{equation}
\label{eq:momentum}
\rho(\mathbf{x}, t)\mathbf{u}(\mathbf{x}, t) = \sum_{i=0}^{Q-1} f_i(\mathbf{x}, t)\mathbf{c}_{i} + \frac{1}{2}\mathbf{g}\Delta t,
\end{equation}
\begin{equation}
\label{eq:tensor}
\mathbf{\Pi}(\mathbf{x}, t) = \sum_{i=0}^{Q-1} f_i(\mathbf{x}, t)\mathbf{c}_{i}\mathbf{c}_{i}.
\end{equation}
Note that the momentum flux, $\mathbf{\Pi}(\mathbf{x}, t)$, can be presented by the sum of the equilibrium and non-equilibrium parts, $\mathbf{\Pi}(\mathbf{x}, t)  = \mathbf{\Pi}_{\text{eq}} (\mathbf{x}, t)  + \mathbf{\Pi}_{\text{neq}}(\mathbf{x}, t)$.

\subsection{Smagorinsky subgrid-scale modeling}\label{sec:method-sgs}
In this part, the lattice-Boltzmann-based Smagrinsky SGS large-eddy simulations techniques are summarized. In LBM framework, the effective viscosity $\nu _{\mathrm{eff}}$ (\cite{smagorinsky1963general, hou1994lattice, koda2015lattice}) is modeled as the sum of the molecular viscosity $\nu_0$ and the turbulent viscosity $\nu_t$:
\begin{equation}
\nu _{\mathrm{eff}}=\nu _0+\nu_t, \hspace{.6in} \nu_t = C_{\mathrm{smag}}\Delta ^2\left |\bar{\mathbf{S}}\right |,
\label{smogrinsky_model}
\end{equation}
where $\left |\mathbf{\bar{S}} \right |$ is the filtered strain rate tensor:
\begin{equation}
\label{eq: strain}
\left |\bar{\mathbf{S}}\right | = \frac{-\tau_i \rho \Delta x^2/\Delta t + \sqrt{(\tau_i \rho)^2  \Delta x^4/\Delta t^2 + 18\sqrt{2}\rho C_{\mathrm{smag}} \delta^2 Q^{1/2}}}{6 \rho C_{\mathrm{smag}} \Delta^2},
\end{equation}
where $C_{\mathrm{smag}}$ is the Smagorinsky constant, $\Delta$ represents the filter size, and $\tau_i$ is the relaxation time for the moment-space collision. $Q^{1/2}$ is $Q^{1/2} = \sqrt{\mathbf{\Pi}_{\mathrm{neq}}:\mathbf{\Pi}_{\mathrm{neq}}}$, with $\mathbf{\Pi}_{\mathrm{neq}}$ is the non-equilibrium part of the momentum flux tensor shown in~\cref{eq:tensor}. With help of \cref{eq:nu}, the total relaxation time $\tau _{\mathrm{eff}}$ is obtained:
\begin{equation}
\label{eq:tau_eff}
    \tau^{\mathrm{eff}}_i=\frac{\tau_i}{2}+\frac{\sqrt{(\tau_i\rho \Delta x/\Delta t)^2  + 18\sqrt{2} C_{\mathrm{smag}} Q^{1/2}}}{2\rho c}.
\end{equation}
Finally, $\tau^{\mathrm{eff}}_i$ is replaced into related MRT collision operator relaxation $\tau_i$ to enclosure the lattice-Boltzmann-based LES system.

\subsection{Near-wall modeling for LBM}\label{sec:force-method}
There have been various approaches in the LBM framework to model the near-wall field with a wall model. Most of them are based on reconstruction of near-wall populations to preserve the velocity and density fields or bounce-back approach to preserve the target velocity (\cite{malaspinas2014wall, haussmann2019large, maeyama2020unsteady, pasquali2020near, wilhelm2021new, asmuth2021wall}).  \cite{malaspinas2014wall} also coupled a RANS solver at the near-wall region with LBM, however, it is proven to be more time consuming than a monolithic LBM
method. In the present work, we focus on developing a wall model that based on forces in the near-wall region.
\begin{figure}
\centering
\includegraphics[width=0.5\textwidth]{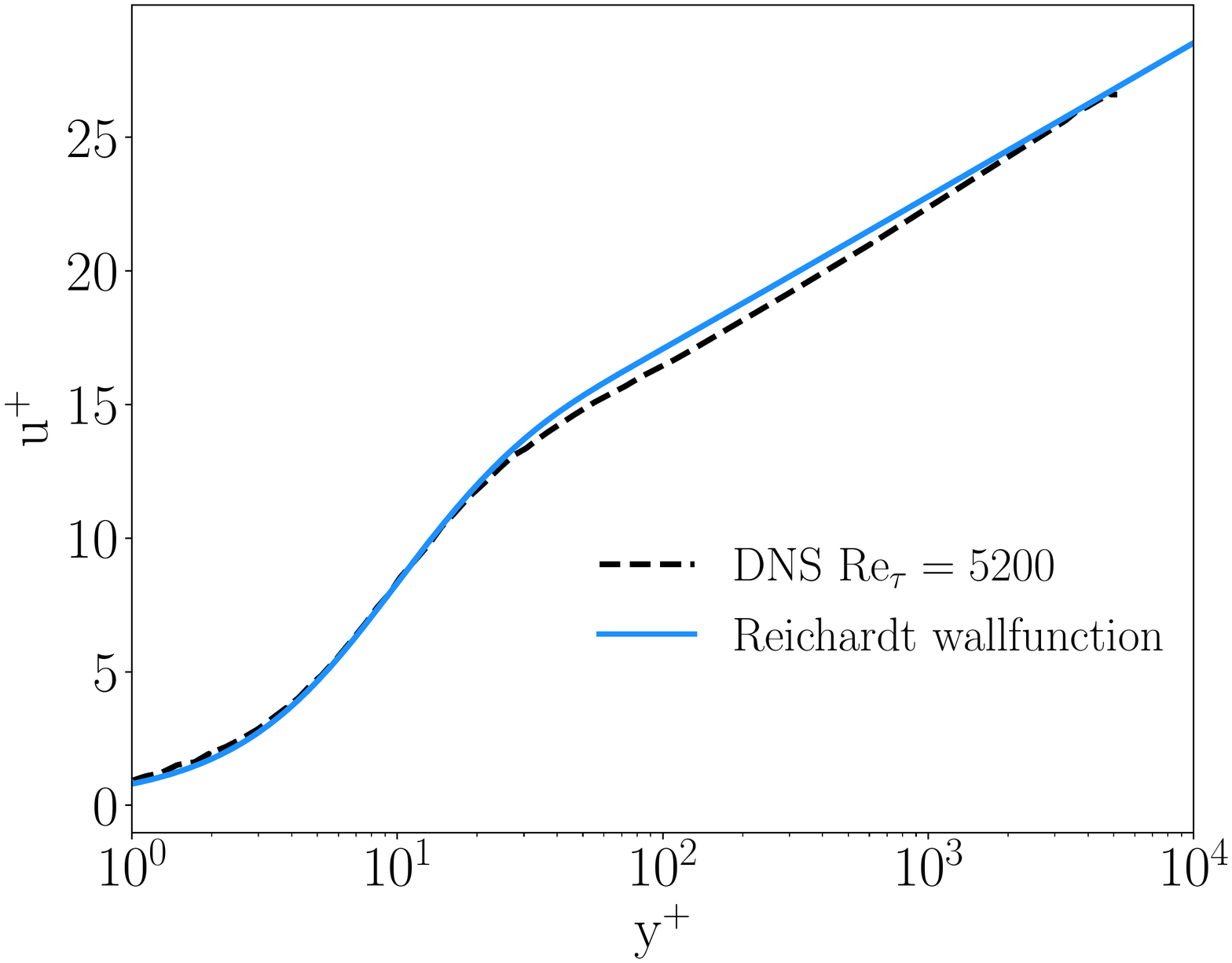}
\caption{Reichardt's wall law  from $y^+ = 1$ to $y^+=10000$.}\label{fig:reichardt} 
\end{figure}

\begin{figure}
\centering
\includegraphics[width=0.38\textwidth]{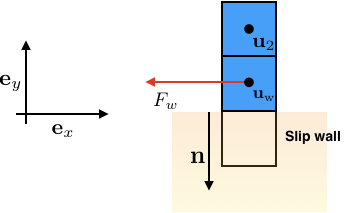}
\caption{Sketch on the first layer near the wall}
\label{fig:force_illustration} 
\end{figure}

In this paper, the Reichardt wall law (\cite{reichardt1951vollstandige}) will be used instead of the one from Musker (\cite{musker1979explicit}). The Reichardt wall law shown in Fig. \ref{fig:reichardt} is defined as
\begin{equation}
\label{eq:reichard}
u^+ = 2.5 \mathrm{ln}(1 + 0.4 y^+) + 7.8 (1 - e^{-y^+ / 11}) -  y^+ e^{-y^+ / 3},
\end{equation}
where $u^+$ and $y^+$ are the dimensionless unit defined as
\begin{equation}
\label{eq:y_plus}
 u^{+}=\frac{\langle  u  \rangle}{u_{\tau}}, \hspace{.4in} y^{+}=y u_\tau/{\nu},
\end{equation}

where $u_{\tau}$ is the shear velocity $u_{\tau} = \sqrt{\tau_{w}/\rho}$, $\left\langle \cdot \right\rangle$ denote ensemble average over space or time, and $\tau_{w}$ is the wall shear stress, $u$ is the streamwise velocity. Figure \ref{fig:force_illustration} illustrates the use of a slip wall boundary condition for the wall treatment. To generalize the wall model, a base vector $\mathbf{e}_x$ is first needed to project the near-wall velocity $\mathbf{u}_{\text{w}}$ on the wall-parallel direction:
\begin{equation}
\label{eq:base}
\mathbf{e}_x=\frac{\mathbf{u}_2- (\mathbf{u}_2\cdot \mathbf{n})\mathbf{n}}{\left \| \mathbf{u}_2- (\mathbf{u}_2\cdot \mathbf{n})\mathbf{n} \right \|},
\end{equation}
where $\mathbf{u}_2$ is the velocity in the second cell from the wall and $\mathbf{n}$ is the wall-normal vector. Then, we project the velocity in the first cell near the wall, $\mathbf{u}_{\text{w}}$, to obtain the scalar streamwise velocity, $\hat{u}_{\text{w}}$, which is defined as
\begin{equation}
\label{eq:uhat}
\hat{u}_{\text{w}}=\mathbf{u}_{\text{w}}\cdot\mathbf{e}_x.
\end{equation}
Next, the aim is to compute the friction velocity $u_{\tau}$ by solving for $u_{\tau}$ in.
\cref{eq:reichard}
\begin{equation}
u_{\tau}(\mathbf{x}, t)=u^+(y_{\perp}(\mathbf{x}, t), \hat{u}_{\text{w}}(\mathbf{x}, t), u_{\tau}(\mathbf{x}, t)).
\label{eq:u_tau_implicit}
\end{equation}
By using the Newton method, we update the friction velocity locally. Note that, instead of using a plane averaged friction velocity (\cite{haussmann2019large, asmuth2021wall}), this work uses the local value to make the algorithm more generalized. The wall shear stress can be estimated by $\tau_{\text{w}}(\mathbf{x}, t) = u_{\tau}^2(\mathbf{x}, t) \rho(\mathbf{x}, t)$.
Finally, the force near the wall is defined as
\begin{equation}
\label{eq:force_wm}
F_{\text{w}}(\mathbf{x}, t) = - \tau_{\text{w}}(\mathbf{x}, t) A,
\end{equation}
where $F$ is the shear force acting on the wall. A notable advantage of the force-based method is that it does not require any reconstruction of populations to find the target velocity and density near the wall, making it easier to implement compared to other wall modeling methods.

\section{Synthetic turbulence generator for the wall-modeled lattice Boltzmann method}\label{sec:num-stg}
\subsection{Synthetic turbulence generator formulation in LBM framework}\label{sec:method-stg}
In this study, a synthetic turbulence generator (STG) is positioned at the inlet of a channel flow. It needs velocity field from a $k-\omega$ RANS simulation (\cite{wilcox:88}). The total velocity $\mathbf{u}_{\text{in}}(\mathbf{x}, t)$ at the inlet is given by
\begin{equation}
\label{eq:u_couple}
\mathbf{u}_{\text{in}}(\mathbf{x}, t) = \mathbf{u}_{\text{RANS}}(\mathbf{x}) + \mathbf{u}'(\mathbf{x}, t),
\end{equation}
where $\mathbf{u}_{\text{RANS}}$ is the velocity vector obtained from a 1D RANS simulation. The STG generates the velocity fluctuations $\mathbf{u}'(\mathbf{x}, t)$ with at the cell $\mathbf{x}$ at time $t$:
\begin{equation}
\label{eq:v_fluc}
\mathbf{u}'(\mathbf{x}, t)=a_{\alpha\beta}\mathbf{v}'(\mathbf{x}, t).
\end{equation}
The time-averaged velocity fluctuation is zero, i.e. $\langle \mathbf{u}'(\mathbf{x}, t)\rangle = 0$. The Cholesky decomposition of the Reynolds stress tensor reads:
\begin{equation}
\label{eq:cholesky}
\left \{  a_{\alpha\beta}\right \}  = 
\begin{pmatrix}
\sqrt{R_{11}} & 0 & 0\\ 
R_{21}/a_{11} & \sqrt{R_{22}-a_{21}^2} & 0\\ 
R_{31}/a_{11} & (R_{32}-a_{21}a_{31})/a_{22} & \sqrt{R_{33}-a_{31}^2-a_{32}^2}
\end{pmatrix},
\end{equation}
where $R_{\alpha\beta} = \left\langle u'_\alpha u'_\beta \right\rangle$ is taken from Reynolds stress tensor using EARSM (\cite{wallin:johansson:00}) when post-processing the 1D RANS data.
$\mathbf{v}'(\mathbf{x}, t)$ in \cref{eq:v_fluc} is imposed by $N$ Fourier modes given by
\begin{equation}
\label{eq:aux_vec} 
\mathbf{v}'(\mathbf{x}, t)  = \sqrt{6} \sum_{n=1}^{N}\sqrt{q^n}\left [ \mathbf{\sigma}^n \mathrm{cos}\left ( k^n\mathbf{d}^n\cdot\mathbf{x}' + \phi^n  \right ) \right ],
\end{equation}
where $q^n$ is the amplitude of a modified von Karman spectrum, $n$ is the mode number, $k^n$ is the amplitude of the mode direction vectors $\mathbf{d}^n$ with $\mathbf{\sigma}^n \cdot \mathbf{d}^n = 0$, $\phi^n$ is the random mode phase that is uniformly distributed in the interval of $[0, 2\pi)$.  A detailed description is found in \cite{shur2014synthetic} and \cite{xue2022synthetic}. The distribution function at the inlet of the channel flow can be defined as the sum of equilibrium part and the non-equilibrium part
\begin{equation}
\label{eq:f_i}
f^{\text{in}}_i(\mathbf{x}, t) = f^{\text{in(eq)}}_i(\mathbf{x}, t) + f^{\text{in(neq)}}_i(\mathbf{x}, t),
\end{equation}
where $f^{\text{in(eq)}}_i(\mathbf{x}, t)$ is the equilibrium part of the inlet distribution function which can be calculated by
\begin{equation}
\label{eq:bc_eq}
f^{\text{in(eq)}}_i\left(\textbf{x},t\right) =  w_{i}\rho_{\text{in}}(\textbf{x},t) \bigg[1+\frac{\textbf{c}_i\cdot\mathbf{u}_{\text{in}}(\mathbf{x}, t)}{c_s^2}+\frac{\left[\textbf{c}_i\cdot\mathbf{u}_{\text{in}}(\mathbf{x}, t)\right]^2}{2c_s^4}  - \frac{\left[\mathbf{u}_{\text{in}}(\mathbf{x}, t)\cdot\mathbf{u}_{\text{in}}(\mathbf{x}, t)\right]}{2c_s^2}\bigg] \;.
\end{equation}
Following the regularized scheme~\cite{latt2008straight}, the non-equilibrium part of the inlet distribution function in~\cref{eq:f_i} is obtained via
\begin{equation}
\label{eq:f_i_neq}
f^{\text{in(neq)}}_i(\mathbf{x}, t)\approx \frac{w_i}{c_s^4} \mathbf{Q}_i:\mathbf{\Pi}^{\text{in}}_{\text{neq}},
\end{equation}
where $\mathbf{Q}_i = \mathbf{c}_i\mathbf{c}_i - c_s^2\mathbf{I}$ with $\mathbf{I}$ being the identity matrix. $\mathbf{\Pi}^{\text{in}}_{\text{neq}}$ is the non-equilibrium part of the moment flux tensor which is defined as
\begin{equation}
\label{eq:pi_neq}
\mathbf{\Pi}_{\text{neq}} = \sum^{Q-1}_{i=0}\mathbf{Q}_if^{\text{in(neq)}}_i(\mathbf{x}, t).
\end{equation}
The unknown variables in the $i$ th direction at the inlet can be calculated via the known direction following
$\mathbf{Q}_i = \mathbf{\bar{Q}}_{\text{inv}(i)}$, $ f^{\text{in(neq)}}_i(\mathbf{x}, t) = \bar{f}^{\text{in(neq)}}_{\text{inv}(i)}(\mathbf{x}, t)$, where the notation ``inv" denotes the opposite direction of the unknown variable. For the density of the inlet boundary, we follow the idea from ~\cite{zou1997pressure}.
\begin{equation}
\label{eq:density_bc}
\rho_{\text{in}}(\textbf{x},t) = \frac{1}{1+\hat{u}_{\text{in}}(\mathbf{x}, t)}(2\rho_{\bot}(\textbf{x},t) + \rho_{\parallel}(\textbf{x},t)),
\end{equation}
where $\hat{u}_{\text{in}}$ is the cross product with the normal unit vector $\mathbf{n}$ at the boundary $\hat{u}_{\text{in}}=\mathbf{u}^{\text{in}}_{\text{LB}}\cdot \mathbf{n}(\left | \hat{u}_{\text{in}} \right |<0.3c_s)$ and $\mathbf{u}^{\text{in}}_{\text{LB}}$ is the velocity of the lattice Boltzmann domain at the interface. $\rho_{\bot}$ and $\rho_{\parallel}$ are the density calculated by
\begin{equation}
\label{eq:density_bc_sub}
\rho_{\bot}(\textbf{x},t)=\sum_{i\in\left\{ i| \mathbf{c}_i \cdot \mathbf{n}' = 0\right\}}f^{\bot}_i(\mathbf{x}, t),\hspace{.6in}\rho_{\parallel}(\textbf{x},t) = \sum_{ i\in\left\{ i| \mathbf{c}_i \cdot \mathbf{n}' < 0\right\}}f^{\parallel}_i(\mathbf{x}, t),
\end{equation}
where $\mathbf{n}'$ is the normal vector pointing towards the inlet boundary, $f^{\bot}_i$ and $f^{\parallel}_i$ are the probability density functions that point towards the boundary and are parallel to the boundary. 



\subsection{Implementation summary}\label{sec:method-sgs-sum}
Below, we summarize the implementation details of STG in the WMLBM framework.
\begin{algorithm}[H]
   \caption{STG-WMLBM: Implementation of synthetic turbulence generator for the MRT wall-modeled LBM}
   \label{alg:stg-lbm}
   \begin{algorithmic}
    \STATE 1. Obtain the RANS velocity at the inlet $\mathbf{u}_{\text{RANS}}(\mathbf{x})$ in~\cref{eq:u_couple}.
    \STATE 2a. Read saved value from the RANS simulation: $R_{\alpha\beta}$, $k$, $\omega$ field etc.
    \STATE 2b. Compute the Reynolds stresses using EARSM
    \STATE 3. Calculate $a_{\alpha\beta}$ with help of~\cref{eq:cholesky}.\\
    \textbf{for all} $t$ from $0$ to $t_{end}$ \textbf{do}\\
        \hskip1.0em \textbf{for all} cells \textbf{do}\\
            \hskip2.0em \textbf{if} cell $\mathbf{x}$ is at the RANS/LBM inlet \textbf{then}\\
                \STATE \hskip3.0em  4. Calculate $\mathbf{v}'(\mathbf{x}, t)$ from~\cref{eq:aux_vec}.\\
                \STATE \hskip3.0em  5. Compute the fluctuating velocity $\mathbf{u}'(\mathbf{x}, t)$ following~\cref{eq:v_fluc} and~\cref{eq:cholesky} respectively.\\
                \STATE \hskip3.0em  6. Compute boundary density $\rho_{\text{in}}(\mathbf{x}, t)$ thanks to~\cref{eq:density_bc}.\\
                \STATE \hskip3.0em  7. Reconstruct the particle's probability distribution function by combining~\cref{eq:f_i}, \cref{eq:bc_eq} and \cref{eq:f_i_neq}.\\
                \STATE \hskip3.0em  8. Update the LES-LBM relaxation time for MRT framework $t_{\text{eff}}$ in~\cref{eq:tau_eff} and replace in kinematic-viscosity-related relaxation time in~\cref{eq:nu}.\\
            \hskip2.0em \textbf{end if}\\
            
            \hskip2.0em \textbf{if} cell $\mathbf{x}$ is at the wall-function cell \textbf{then}\\
                \STATE \hskip3.0em  9. Compute $\mathbf{u}_2$.\\
                \STATE \hskip3.0em  10. Compute the wall-parallel base vector $\mathbf{e}_x$ with help of~\cref{eq:base}.\\
                \STATE \hskip3.0em  11. Compute the scalar streamwise velocity $\hat{u}_{\text{w}}$ using~\cref{eq:uhat}.\\
                \STATE \hskip3.0em  12. Solve implicit function to obtain $u_{\tau}$ with help of~\cref{eq:u_tau_implicit} and~\cref{eq:reichard}.\\
                \STATE \hskip3.0em  13. Compute $\tau_{\text{w}}$ with help of $\tau_{\text{w}}(\mathbf{x}, t) =u_\tau^2(\mathbf{x}, t) \rho(\mathbf{x}, t)$.\\
                \STATE \hskip3.0em  14. Update force on the cell $F_{\text{w}}(\mathbf{x}, t)$ by using~\cref{eq:force_wm}\\
            \hskip2.0em \textbf{end if}\\

            \STATE \hskip2.0em  15. Apply stream and collide with consideration of forces to update the $f_i(\mathbf{x}, t)$ at each cell~\cref{eq:lbe}\\
        \hskip1.0em\textbf{end for}\\
    \textbf{end for}\\
   \end{algorithmic}
\end{algorithm}


\section{Turbulent channel flow simulations}\label{sec:results}
This section presents the turbulent channel flow simulations with the STG as the inlet at $Re_{\tau} = 1000$, $Re_{\tau} = 2000$, and $Re_{\tau} = 5200$. The present work employs three different resolutions, LBU, in reference to the height of the channel $H$, that is, 
LBU $=H/20$, $H/40$, and $H/60$.
\subsection{Numerical setup}\label{sec:setup}

\begin{figure}
\centering
\includegraphics[width=0.8\textwidth]{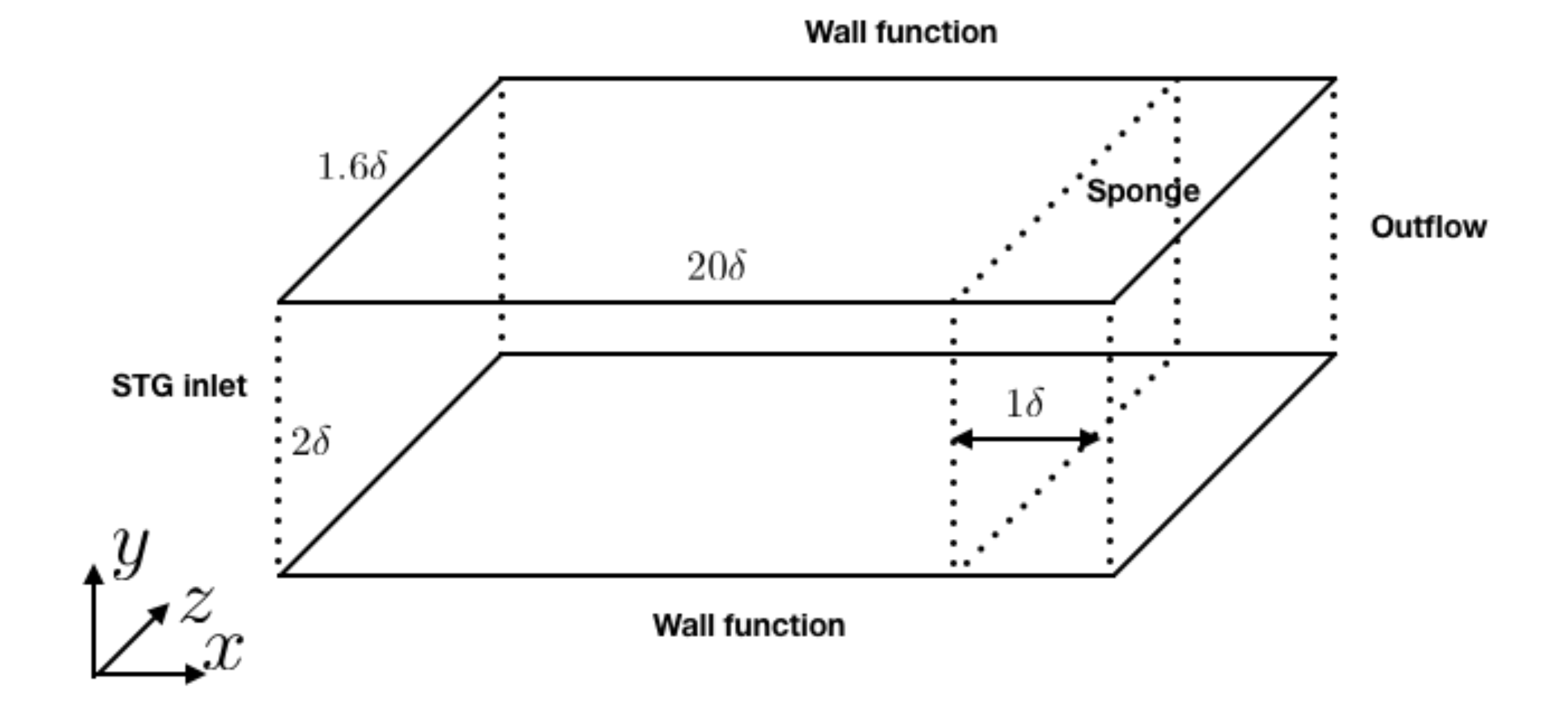}
\caption{Numerical setup of the channel flow simulation} 
\label{fig:sketch_sim}
\end{figure}

The turbulent channel flow simulations use STG at the inlet and a pressure-free in the streamwise direction at the outlet. Periodic boundary conditions are employed in the spanwise direction ($z$) and wall-functions are used at the first cell-layer near the top and bottom ($y$) planes which equipped with the slip boundary condition. To minimize the reflection 
wave's
impact on the flow field, a sponge zone is placed near the outlet (\cite{xue2022synthetic}). The numerical setup for the channel flow is depicted in \cref{fig:sketch_sim}, where the boundary layer thickness ($\delta$) is defined as half of the channel height. The extent of simulation domain is $20\delta \times 2\delta \times 1.6\delta $ in $x$, $y$, and $z$ directions respectively. The sponge layer thickness is set to $1\delta$. The Smagorinsky constant is set to $C_{smag} = 0.01$. The simulations are conducted at different friction Reynolds numbers and resolutions, and run for a total of 10 domain-through times ($10T$). Statistical analysis begins after 2 domain-through time ($2T$).

\begin{figure}
\centering
\includegraphics[width=1\textwidth]{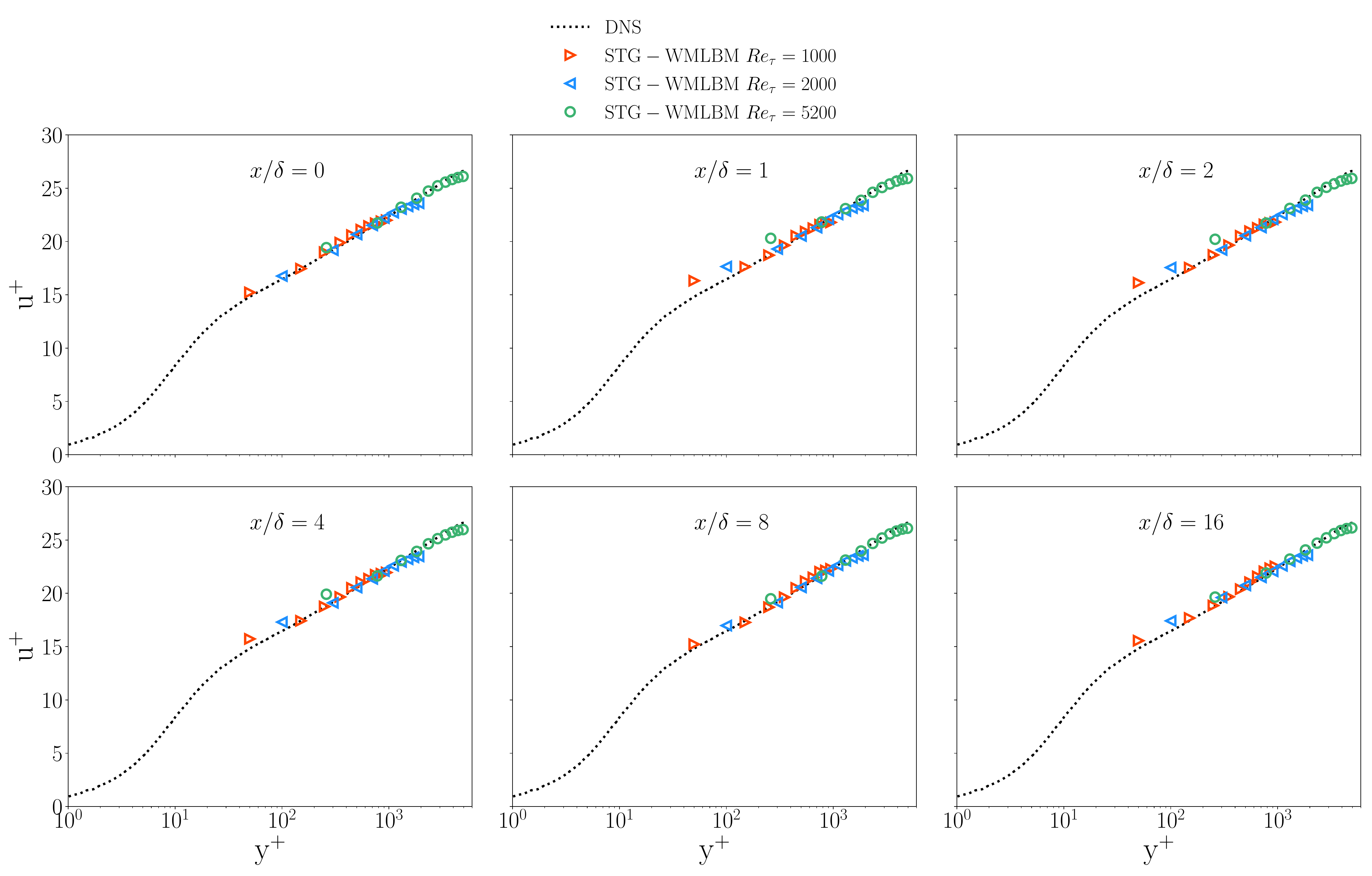}
\caption{$\mathrm{u}^{+}$ as function of $\mathrm{y}^{+}$ at $Re_{\tau} = 1000, 2000, 5200$ for $H=20$ (LBU). } 
\label{fig:u_plus_20}
\end{figure}
\subsection{Results}\label{sec:5.3}
This work presents validation of the STG framework for high friction Reynolds numbers, i.e. $Re_{\tau} = 1000$, $Re_{\tau} = 2000$, and $Re_{\tau} = 5200$. The initial investigation focuses on the resolution with $H=2\delta=20$ LBU. The $y^+$ values at $H=20$ LBU for $Re_{\tau} = 1000$, $2000$, and $5200$ are approximately 50, 100, and 260, respectively. The results, triggered by the STG inlet, are compared with DNS data obtained by \cite{hoyas2006scaling, lee2015direct}.

Figure \ref{fig:u_plus_20} shows mean velocity field of $\mathrm{u}^{+}$ as function of $\mathrm{y}^{+}$. The STG-WMLBM results compare with the DNS data at different friction Reynolds numbers. 
The results of the STG-WMLBM at all three friction Reynolds numbers show good agreement with DNS reference starting from $x/\delta = 0$. While the initial results are promising, further investigations are needed to analyze the Reynolds stresses in order to fully evaluate the effectiveness of the method.

\begin{figure}
\centering
\includegraphics[width=1\textwidth]{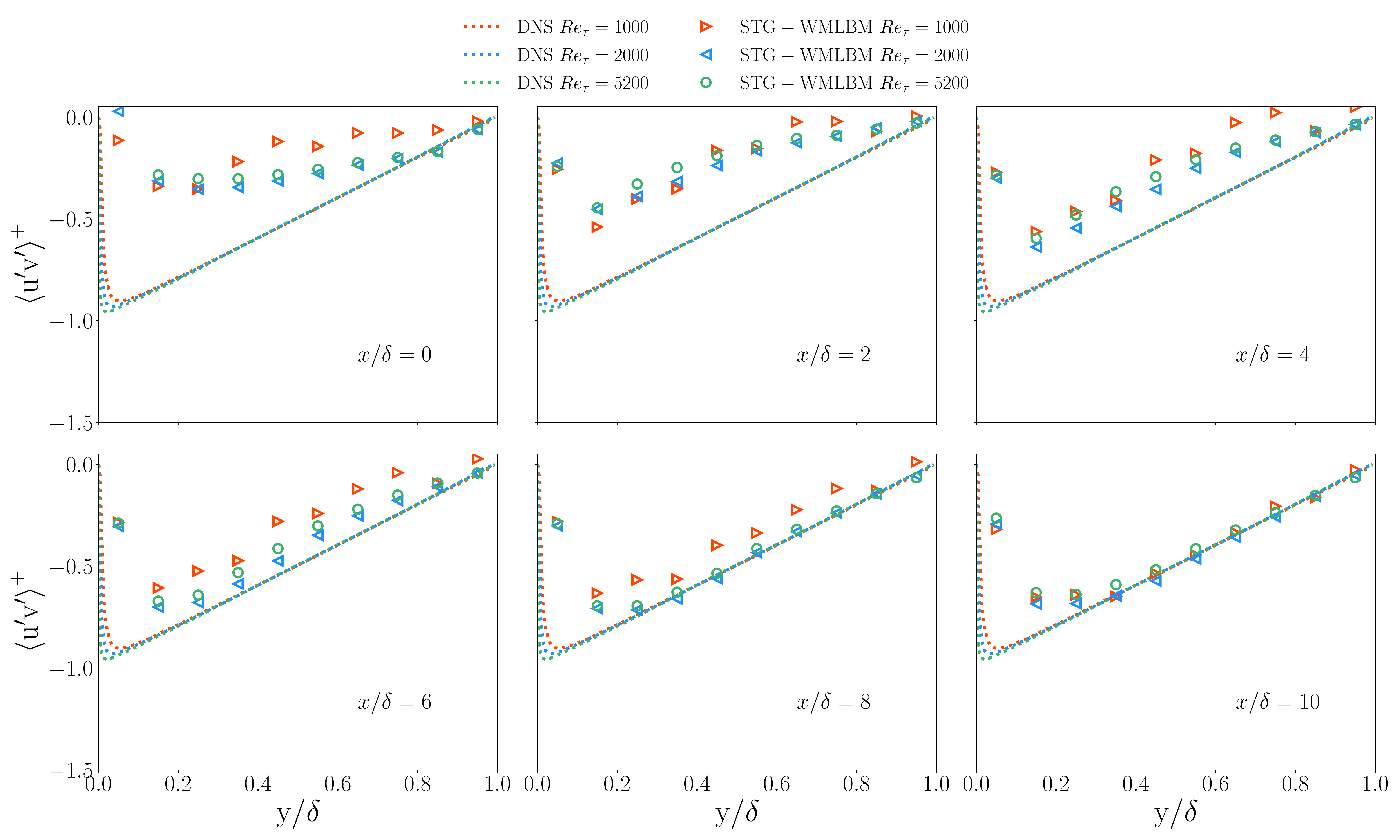}
\caption{$\left \langle \mathrm{u}^\prime\mathrm{v}^\prime \right \rangle^+$ as function of $\mathrm{y/\delta}$ at $Re_{\tau} = 1000, 2000, 5200$ for $H=20$ (LBU).} 
\label{fig:uw_plus_20}
\end{figure}

Figure \ref{fig:uw_plus_20} shows $\left \langle \mathrm{u}^\prime\mathrm{v}^\prime \right \rangle^+$  as function of $\mathrm{y/\delta}$. The STG-WMLBM results compare with the DNS data at different friction Reynolds numbers. 
For $Re_{\tau} = 2000$ and $5200$, the STG-WMLBM results exhibit excellent agreement with the DNS reference data after $x/\delta = 6-8$. Meanwhile, for $Re_{\tau}=1000$, the results converge to the DNS data at approximately $x/\delta = 8-10$. However, discrepancies with the DNS data are observed near the wall, which can be attributed to the poor resolution.
\begin{figure}
\centering
\includegraphics[width=1\textwidth]{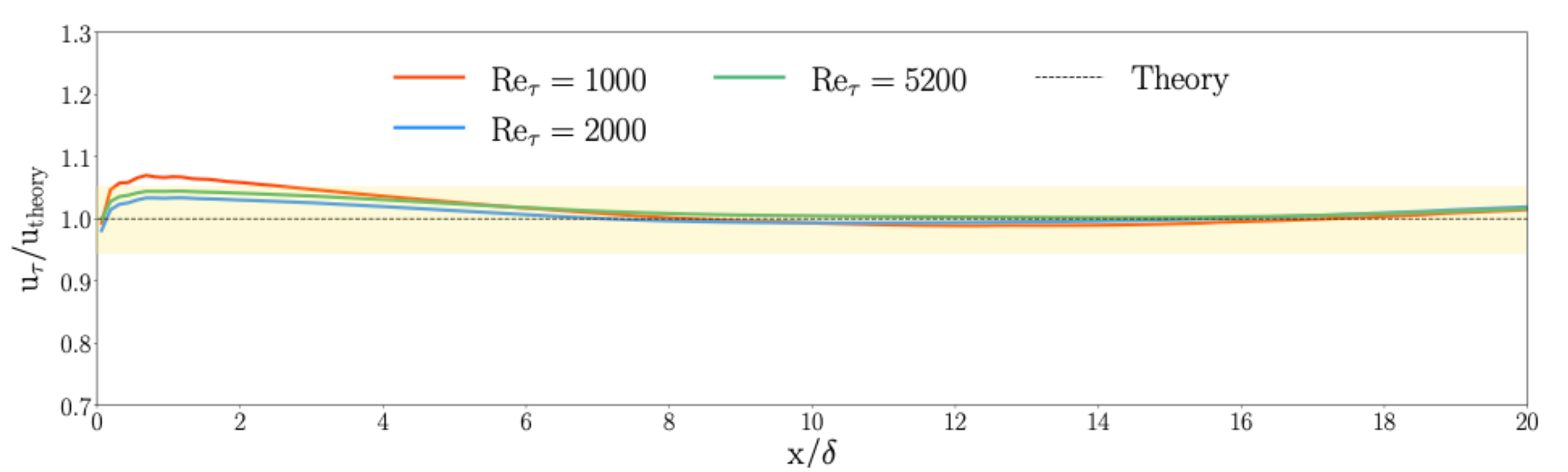}
\caption{Normalised $\mathrm{u}_{\tau}/\mathrm{u}_{\text{theory}}$ as function of $\mathrm{x/\delta}$ at $Re_{\tau} = 1000, 2000, 5200$ for $H=20$ (LBU). } 
\label{fig:u_tau_norm}
\end{figure}

Next, we will present the friction velocity $u_{\tau}/u_{\text{theory}}$ at the three different friction Reynolds numbers $Re_{\tau} = 1000, 2000, 5200$. Figure \ref{fig:u_tau_norm} demonstrates that $u_{\tau}/u_{\text{theory}}$ rapidly converges to the high-accuracy region (shown in yellow, with a relative error of $5\%$) for all three friction Reynolds numbers at $x/\delta = 2$ to $x/\delta = 4$.
\begin{figure}
\centering
\includegraphics[width=1\textwidth]{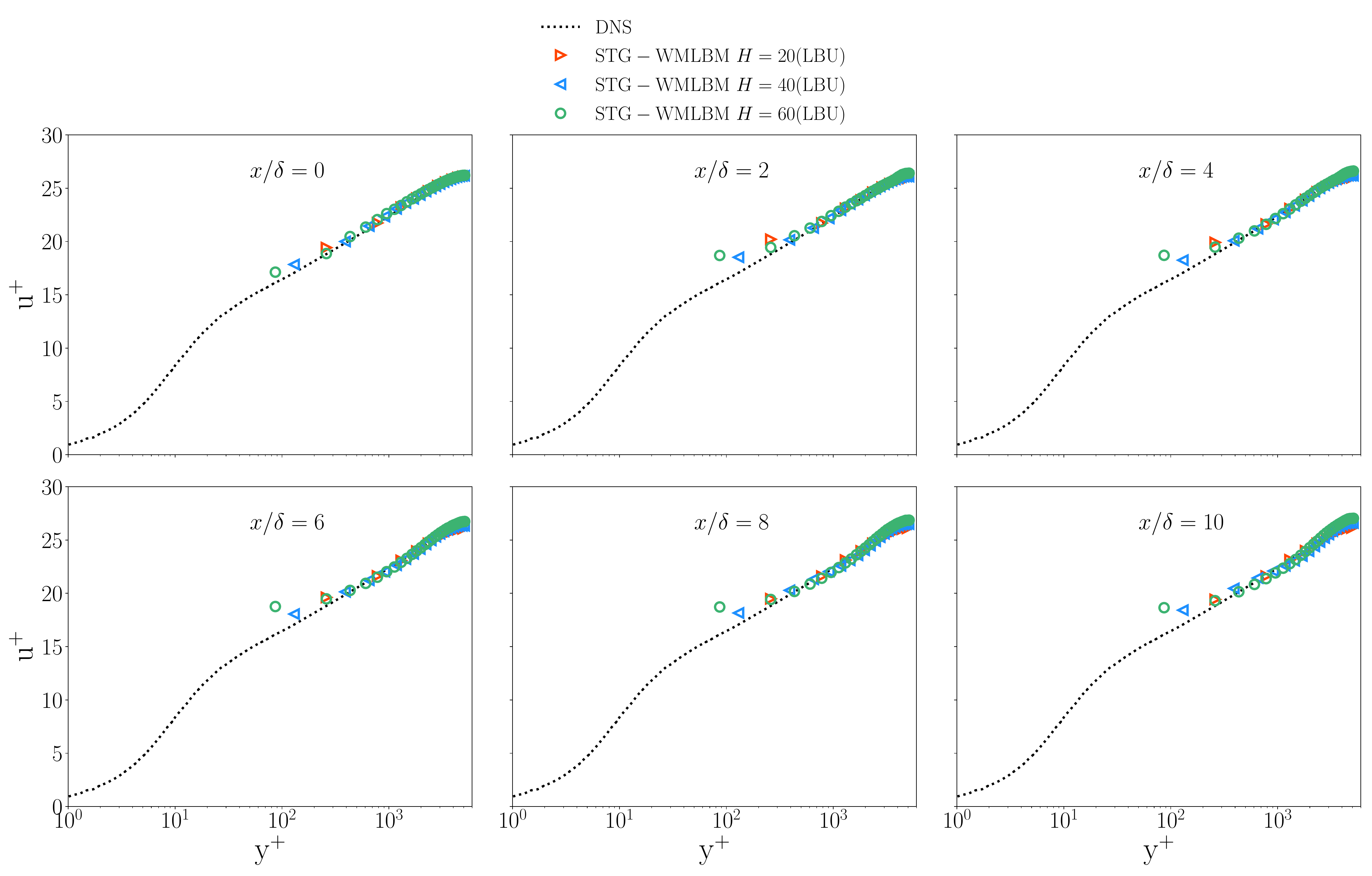}
\caption{$\mathrm{u}^{+}$ as function of $\mathrm{y}^{+}$ at $Re_{\tau} = 5200$ for different resolutions $H=20,40,60$ (LBU).} 
\label{fig:u_plus_5200}
\end{figure}

Further investigation of STG is carried out at $Re_{\tau} = 5200$ for different resolutions of $H = 20, 40, 60$ (LBU), with the first cell $y^+$ equal to $y^+ \approx 260, 130, 86.7$,  respectively. 
Although the first few near-wall cells are slightly off the reference, the STG results match well with DNS further away from the wall, see Fig.~\ref{fig:u_plus_5200}.

\begin{figure}
\centering
\includegraphics[width=0.7\textwidth]{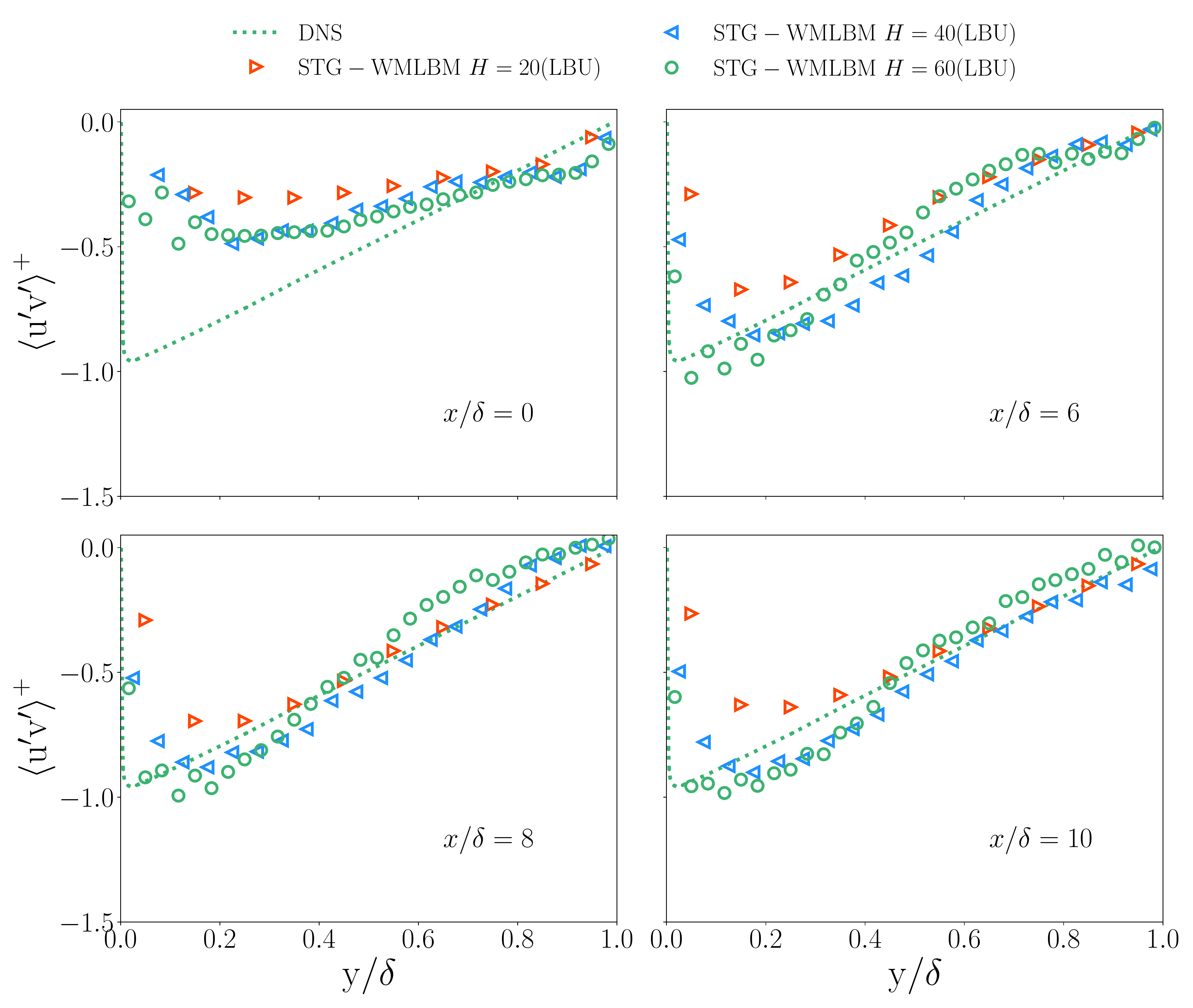}
\caption{$\left \langle \mathrm{u}^\prime\mathrm{v}^\prime \right \rangle^+$ as function of $\mathrm{y/\delta}$ at $Re_{\tau} = 5200$  for different resolutions $H=20,40,60$ (LBU).} 
\label{fig:uv_5200}
\end{figure}
The analysis of $\left \langle \mathrm{u}^\prime\mathrm{v}^\prime \right \rangle^+$ as a function of $\mathrm{y/\delta}$ is carried out at all three resolutions. Figure \ref{fig:uv_5200} displays the results of the STG data with the DNS reference,
The results converge to the DNS reference around $x/\delta = 6$ to $x/\delta = 8$. 

\section{Conclusions}\label{sec:conclusions}
This paper presents a synthetic turbulent generator (STG) model based on the LES-LBM framework for high friction Reynolds number simulations ($Re_{\tau} = 1000, 2000, 5200$). We use wall function based on Reichardt's law (\cite{reichardt1951vollstandige}) in combination with the force-based method which simplifies the implementation. The STG simulations are examined at different resolutions ($H=20, 40, 60$ (LBU)) and compared with the DNS data, showing immediate convergence to the mean velocity field from the inlet of the channel flow. Further analysis of the Reynolds stress indicates that convergence to the DNS data occurs around $x/\delta=6$ to $x/\delta=8$. The presented STG model is computationally efficient and quickly converges to the DNS data even at low resolutions, which is promising for high Reynolds-number applications.


\section{Declaration of interests}
The authors declare that they have no known competing financial interests or personal relationships that could have appeared to influence the work reported in this paper.

\section{Acknowledgements}
The authors kindly acknowledge the funding from Chalmers Transport Area of Advance. The computations and data handling were enabled by resources provided by the Swedish National Infrastructure for Computing (SNIC), partially funded by the Swedish Research Council through grant agreement no. 2018-05973.

\bibliography{prex}
\section{Appendix}
\subsection{D3Q19 MRT matrix and choose of parameters}\label{sec:A.1}
The MRT matrix $\mathbf{M}$ in \cref{eq:collision} is defined as:
\begin{equation}
 \mathbf{M}=\left[
\arraycolsep=1pt
\begin{array}{ccccccccccccccccccc}
 1 &1 & 1 & 1 & 1 & 1 & 1 & 1 & 1 & 1 & 1 & 1 & 1 & 1 & 1 & 1 & 1 & 1 & 1 \\
 8 & 8 & -11 & 8 & 8 & 8 & -11 & 8 & -11 & -30 & -11 & 8 & -11 & 8 & 8 & 8 & -11 & 8 & 8 \\
 1 & 1 & -4 & 1 & 1 & 1 & -4 & 1 & -4 & 12 & -4 & 1 & -4 & 1 & 1 & 1 & -4 & 1 & 1 \\
 0 & -1 & 0 & 1 & 0 & -1 & 0 & 1 & -1 & 0 & 1 & -1 & 0 & 1 & 0 & -1 & 0 & 1 & 0 \\
 0 & -1 & 0 & 1 & 0 & -1 & 0 & 1 & 4 & 0 & -4 & -1 & 0 & 1 & 0 & -1 & 0 & 1 & 0 \\
 -1 & 0 & 0 & 0 & 1 & -1 & -1 & -1 & 0 & 0 & 0 & 1 & 1 & 1 & -1 & 0 & 0 & 0 & 1 \\
 -1 & 0 & 0 & 0 & 1 & -1 & 4 & -1 & 0 & 0 & 0 & 1 & -4 & 1 & -1 & 0 & 0 & 0 & 1 \\
 -1 & -1 & -1 & -1 & -1 & 0 & 0 & 0 & 0 & 0 & 0 & 0 & 0 & 0 & 1 & 1 & 1 & 1 & 1 \\
 -1 & -1 & 4 & -1 & -1 & 0 & 0 & 0 & 0 & 0 & 0 & 0 & 0 & 0 & 1 & 1 & -4 & 1 & 1 \\
 -2 & 1 & -1 & 1 & -2 & 1 & -1 & 1 & 2 & 0 & 2 & 1 & -1 & 1 & -2 & 1 & -1 & 1 & -2 \\
 -2 & 1 & 2 & 1 & -2 & 1 & 2 & 1 & -4 & 0 & -4 & 1 & 2 & 1 & -2 & 1 & 2 & 1 & -2 \\
 0 & -1 & -1 & -1 & 0 & 1 & 1 & 1 & 0 & 0 & 0 & 1 & 1 & 1 & 0 & -1 & -1 & -1 & 0 \\
 0 & -1 & 2 & -1 & 0 & 1 & -2 & 1 & 0 & 0 & 0 & 1 & -2 & 1 & 0 & -1 & 2 & -1 & 0 \\
 0 & 0 & 0 & 0 & 0 & 1 & 0 & -1 & 0 & 0 & 0 & -1 & 0 & 1 & 0 & 0 & 0 & 0 & 0 \\
 1 & 0 & 0 & 0 & -1 & 0 & 0 & 0 & 0 & 0 & 0 & 0 & 0 & 0 & -1 & 0 & 0 & 0 & 1 \\
 0 & 1 & 0 & -1 & 0 & 0 & 0 & 0 & 0 & 0 & 0 & 0 & 0 & 0 & 0 & -1 & 0 & 1 & 0 \\
 0 & 1 & 0 & -1 & 0 & -1 & 0 & 1 & 0 & 0 & 0 & -1 & 0 & 1 & 0 & 1 & 0 & -1 & 0 \\
 -1 & 0 & 0 & 0 & 1 & 1 & 0 & 1 & 0 & 0 & 0 & -1 & 0 & -1 & -1 & 0 & 0 & 0 & 1 \\
 1 & -1 & 0 & -1 & 1 & 0 & 0 & 0 & 0 & 0 & 0 & 0 & 0 & 0 & -1 & 1 & 0 & 1 & -1
\end{array}
\right].
\end{equation}
The relaxation parameters that are not determined by the viscosity are set to:
\begin{align}
\label{eq:freeparams}
\omega_{0} &= \omega_{3}=\omega_{5}=\omega_{7}=\omega_{1}=1.0\\
\omega_1 &= 1.19,\\ 
\omega_2 &= \omega_{10} = \omega_{12} = 1.6\\
\omega_{4} &= \omega_{6} = \omega_{8}=1.2\\
\omega_{16} &= \omega_{17} = \omega_{18}=1.98
\end{align}
\end{document}